\documentclass{llncs}
\usepackage{latexsym}
\usepackage{amsmath,amssymb,eufrak}
\usepackage{epsfig}
\usepackage{pifont}
\usepackage{color}
\usepackage{url}

\def\eqdef{\stackrel{\rm def}{=} \;}
\def\infeq{\mathbin{|\kern -.47em\approx}}
\def\notinfeq{\mathbin{|\kern -.47em\approx\kern-.9em{/}}}

\newcommand{\tuple}[1]{ \langle #1 \rangle }
\newcommand{\V}[1]{\mathbf{#1}}
\def\qed{\hfill$\Box$}

\def\A{\mathbf{A}}
\def\B{\mathbf{B}}
\def\HT{\mathbf{HT}}
\def\HTtup{\tuple{H,T}}
\title{A Logical Characterisation of Ordered Disjunction}

\author{Pedro Cabalar\thanks{This research was partially supported by Spanish MEC project TIN-2009-14562-C05-04 and Xunta de Galicia project INCITE08-PXIB105159PR.}}
\institute{Department of Computer Science,\\
Corunna University (Corunna, Spain),\\
\texttt{cabalar@udc.es}\\[15pt]
}
\begin{document}
\maketitle

\begin{abstract}
In this paper we consider a logical treatment for the ordered disjunction operator $\times$ introduced by Brewka, Niemel\"a and Syrj\"anen in their Logic Programs with Ordered Disjunctions (LPOD). LPODs are used to represent preferences in logic programming under the answer set semantics. Their semantics is defined by first translating the LPOD into a set of normal programs (called split programs) and then imposing a preference relation among the answer sets of these split programs. We concentrate on the first step and show how a suitable translation of the ordered disjunction as a derived operator into the logic of Here-and-There allows capturing the answer sets of the split programs in a direct way. We use this characterisation not only for providing an alternative implementation for LPODs, but also for  checking several properties (under strongly equivalent transformations) of the $\times$ operator, like for instance, its distributivity with respect to conjunction or regular disjunction. We also make a comparison to an extension proposed by K\"arger, Lopes, Olmedilla and Polleres, that combines $\times$ with regular disjunction.
\end{abstract}

\section{Introduction}

Based on the answer set (or stable model) semantics~\cite{GL88} for logic programs, Answer Set Programming (ASP)~\cite{MT99,Nie99} has become a successful paradigm for declarative problem solving. Typically, a logic program in ASP is used to encode some constraint-based problem in such a way that the answer sets of the program correspond to the problem solutions. In many practical scenarios, however, the set of feasible solutions is considerably large and the main problem, from the knowledge representation viewpoint, is not specifying them but selecting the most preferred ones under certain criteria instead. Although different approaches for representing preferences in ASP have been proposed (see~\cite{DSTW04} for a survey), one that has recently received much attention is the formalism of \emph{Logic Programs with Ordered Disjunction} (LPOD)~\cite{BNS04}, probably due to its simplicity and expressiveness. This approach essentially consists in introducing a new operator `$\times$' standing for \emph{ordered disjunction} (with its corresponding semantics in terms of answer sets), plus several ordering relations for selecting preferred models among the obtained answer sets. LPODs have been applied, for instance, in Game Theory~\cite{FMB04}, for implementing policy languages with preferences~\cite{MMP05,BMP05}, or in planning and argumentation scenarios~\cite{ZONSS05}, and they have been further investigated in~\cite{FTW08} for studying strongly equivalent transformations and in~\cite{KLOP08} for introducing an extension called \emph{disjunctive} LPOD (DLPOD) that combines ordered and regular disjunctions. Other ASP extensions like CR-Prolog, have also incorporated the use of ordered disjunctions~\cite{BM03}. The semantics of an LPOD is defined in two steps. First, the program with ordered disjunctions is translated into a set of normal programs, called \emph{split programs}, whose answer sets become the potential solutions. In a second step, one of three possible preference relations is imposed among the answer sets of these split programs. These answer sets of the split programs can also be captured by reduct transformations, like the originally introduced in~\cite{BNS04} or the one later proposed in~\cite{FTW08}.

In this paper we concentrate on the first step, that is, in the definition of potential answer sets for LPODs, and show that they can be directly captured by a suitable definition of the ordered disjunction connective `$\times$' as a derived operator in the logic of \emph{Here-and-There} ($\HT$)~\cite{Hey30}, so that LPODs can be seen as nothing else but regular theories inside the nonmonotonic formalism of  \emph{Equilibrium Logic}~\cite{Pea97} (the nonmonotonic version of $\HT$). Equilibrium Logic has been extensively studied in ASP, as it yields a logical characterisation for the answer set semantics, capturing concepts such as the strong equivalence of programs~\cite{LPV01} and providing a means to generalise all previous extensions to the the most general syntax of arbitrary propositional~\cite{Fer05} and first order~\cite{FLL07} theories. Our logical characterisation of $\times$ allows not only an alternative method for translating LPODs into regular logic programs, but enables the study of `$\times$' as a logical connective, so we can analyse its main properties like its behaviour with respect to distributivity or nesting with other connectives.

The rest of the paper is organised as follows. The next section contains an overview of the basic definitions of Equilibrium Logic, ASP and LPODs. Section~\ref{sec:od} introduces the characterisation of ordered disjunction and studies some of its main properties, including the correspondence to the original definition of answer sets for LPODs. Section~\ref{sec:dlpod} contains a comparison to DLPODs and finally, Section~\ref{sec:conc} concludes the paper.

\section{Preliminaries}

We recall the basic definitions of the propositional\footnote{For simplicity sake, we will focus here on ordered disjunction inside programs without variables.} logic of $\HT$ and  Equilibrium Logic. The syntax is the same as in classical propositional logic: a \emph{well formed formula} results from combining atoms in a finite set $\Sigma$ (called the \emph{signature}) with the usual operators $\rightarrow, \wedge, \vee, \bot$ and parentheses. We assume the standard precedence among binary operators, that is, $\wedge \prec \vee \prec \ \rightarrow$. The formulas $\neg F$, $\top$, $F \leftarrow G$ and $F \leftrightarrow G$ are abbreviations that respectively stand for $F \rightarrow \bot$, $\bot \rightarrow \bot$, $G \rightarrow F$ and $(F \rightarrow G) \wedge (F \leftarrow G)$. As usual, a \emph{literal} is an atom $p$ or its negation $\neg p$.

Given a set of atoms $I$ and a formula $F$, we write $I \models F$ to represent classical satisfaction. The semantics of $\HT$ starts from defining an \emph{interpretation} as a pair $\HTtup$ of sets of atoms (standing for ``here'' and ``there'') where $H\subseteq T$. We say that an interpretation $\HTtup$ \emph{satisfies} a formula $F$, by abuse of notation written  $\HTtup \models F$ as in classical logic, when one of the following recursive conditions hold:
\begin{enumerate}
\item $\HTtup \models p$ if $p \in H$, for any atom $p$.
\item $\HTtup \not\models \bot$.
\item $\HTtup \models F \wedge G$ if $\HTtup \models F$ and $\HTtup \models G$.
\item $\HTtup \models F \vee G$ if $\HTtup \models F$ or $\HTtup \models G$.
\item\label{itarrow} $\HTtup \models F \rightarrow G$ if both (i) $T \models F \rightarrow G$; and (ii) $\HTtup \not\models F$ or $\HTtup \models G$.
\end{enumerate}
\noindent Ambiguity in the use of $\models$ is removed depending on the interpretation we use in the left hand. Thus, note that in condition (i) of line~\ref{itarrow}, we are actually referring to classical satisfaction. Some useful well-known properties relating $\HT$ and classical satisfaction are mentioned below.

\begin{proposition}
For any interpretation $\HTtup$ and any formula $F$:
\begin{enumerate}
\item $\HTtup \models F$ implies $T \models F$.
\item $\tuple{T,T} \models F$ iff $T \models F$.
\item $\HTtup \models \neg F$ iff $T \models \neg F$.\qed
\end{enumerate}
\end{proposition}

By $\equiv_c$ we understand equivalence in classical logic. A \emph{theory} is a set of formulas. As usual, an interpretation $\HTtup$ is said to be a model of a theory $H$, also written $\HTtup \models H$, when $\HTtup$ satisfies all the formulas in $H$.

The following are some $\HT$ valid equivalences:
\begin{eqnarray}
\neg (F \vee G) & \leftrightarrow & \neg F \wedge \neg G \label{demorgan1}\\
\neg (F \wedge G) & \leftrightarrow & \neg F \vee \neg G \label{demorgan2}\\
F \wedge (G \vee H) & \leftrightarrow & (F \wedge G) \vee (F \wedge H) \label{distrib1}\\
F \vee (G \wedge H) & \leftrightarrow & (F \vee G) \wedge (F \vee H) \label{distrib2}\\
F \wedge \neg F & \leftrightarrow & \bot \label{f:bot}\\
(F \rightarrow (G \rightarrow H)) & \leftrightarrow & (F \wedge G \rightarrow H)\label{f:nestarrow}
\end{eqnarray}
\noindent For instance, $\HT$ satisfies De Morgan's laws \eqref{demorgan1}, \eqref{demorgan2}, and distributivity \eqref{distrib1}, \eqref{distrib2}.

\begin{definition}[Equilibrium model]
An $\HT$ interpretation $\tuple{T,T}$ is an \emph{equilibrium model} of a theory $H$ if $\tuple{T,T} \models H$ and there is no $H \subset T$ such that $\HTtup \models H$.\qed
\end{definition}
\emph{Equilibrium Logic} is the (nonmonotonic) logic induced by equilibrium models. An interesting concept for nonmonotonic reasoning is the idea of strong equivalence. Two theories $H_1$ and $H_2$ are said to be \emph{strongly equivalent}, written $H_1 \equiv_s H_2$, if $H_1 \cup \Delta$ and $H_2 \cup \Delta$ have the same equilibrium models, for any theory $\Delta$.

\begin{theorem}[From~\cite{LPV01}]
Two theories $H_1, H_2$ are strongly equivalent iff they are equivalent in $\HT$.\qed
\end{theorem}

\subsection{Logic programs}

We begin introducing some preliminary notation that will be useful later. Let $\A$ be a (possibly empty) list of (possibly repeated) formulas. We write $|\V{A}|$ to stand for the length of $\V{A}$. For any $k \in \{1,\dots,|\A|\}$, by $\A[k]$ we mean the $k$-th expression in $\A$ and by $\A[1..k]$, the prefix of $\A$ of length $k$, that is, $\A[1] \dots \A[k]$. For a binary operator $\odot \in \{\vee, \wedge,\times\}$, by $(\odot \A)$ we mean the formula resulting from the repeated application of $\odot$ to all formulas in $\A$ in the same ordering.  As an example, given the sequence of atoms $\A=(a,b,c,d,e)$, the expression $(\times \A[1..3])$ represents the formula $a \times b \times c$.  We write $\neg \A$ to stand for the sequence of formulas $\neg \A[1] \dots \neg \A[k]$ being $k=|\A|$. An empty conjunction is understood as $\top$ whereas an empty disjunction (both ordered or regular) is understood as $\bot$. The concatenation of two lists of formulas, $\A$ and $\B$, is simply written as $\A \B$.

A \emph{logic program} is a set of rules of the form:
\begin{eqnarray}
(\vee \A) \vee (\vee \neg \A') \leftarrow (\wedge \B) \wedge (\wedge \neg \B') \label{lprule}
\end{eqnarray}
\noindent where $\A,\A',\B$ and $\B'$ are lists of atoms. We respectively call \emph{head} and \emph{body} to the consequent and antecedent of the implication above. A rule with an empty head $\bot$ (that is, $|\A|+|\A'|=0$) is called a \emph{constraint}. A rule with an empty body $\top$ (that is, $|\B|+|\B'|=0$) is called a \emph{fact}, and we usually write the head $F$ instead of $F \leftarrow \top$. A rule is said to be \emph{normal} when $|\A|=1$ and $|\A'|=0$. A rule is \emph{positive} when $|\A'|=|\B'|=0$. We extend the use of these adjectives to a program, meaning that all its rules are of the same kind.

Answer sets of a program $P$ are defined in terms of the classical Gelfond-Lifscthiz's reduct~\cite{GL88}, that is extended\footnote{In fact, \cite{Fer05} introduced a different, more general reduct that allows defining answer sets for arbitrary theories, which happen to coincide with equilibrium models. We use here the more restricted, traditional version for comparison purposes.} as follows for the syntactic case we are considering (disjunctive heads with default negation~\cite{IS98}). The \emph{reduct} of a program $P$ with respect to a set of atoms $I$, written $P^I$, consists of a rule like $(\vee \A) \leftarrow (\wedge \B)$ per each rule in $P$ of the form \eqref{lprule} that satisfies $I \models (\wedge \A') \wedge (\wedge \neg \B')$. We say that a set of atoms $I$ is an \emph{answer set} of a program $P$ if $I$ is a minimal model of $P^I$.

\begin{theorem}[From~\cite{LPV01}]\label{th:as}
A set of atoms $T$ is an answer set of a program $P$ iff $\tuple{T,T}$ is an equilibrium model of $P$.\qed
\end{theorem}

\subsection{Logic Programs with Ordered Disjunction}

A \emph{logic program with ordered disjunction} (LPOD) is a set of rules of the form:
\begin{eqnarray}
(\times \A) \leftarrow (\wedge \B) \wedge (\wedge \neg \B') \label{lpodrule}
\end{eqnarray}
\noindent where $\A, \B$ and $\B'$ are lists of atoms. We say that a set of atoms $I$ \emph{satisfies} an LPOD rule $r$ like \eqref{lpodrule}, written $I \models r$, when $I \models (\vee \A) \leftarrow (\wedge \B) \wedge (\wedge \neg \B')$ in classical logic.

For each LPOD rule $r$ like \eqref{lpodrule}, we define its $k$-th option, written $r_k$, with $k \in \{1,\dots,|\A|\}$, as the normal rule:
\begin{eqnarray*}
\A[k] \leftarrow (\wedge \B) \wedge (\wedge \neg \B') \wedge (\wedge \neg \A[1..k\!\!-\!\!1]) \end{eqnarray*}

A normal logic program $P'$ is a \emph{split program} of $P$ if it is the result of replacing each LPOD rule $r \in P$ by one of its possible options $r_k$. A set of atoms $I$ is an \emph{answer set} of $P$ if it is an answer set of some split program $P'$ of $P$.

\begin{example}[From~\cite{BNS04}]\label{ex1}
Let $P_{\ref{ex1}}$ be the LPOD:
\begin{eqnarray*}
a \times b \leftarrow \neg c & \hspace{40pt} &
b \times c \leftarrow \neg d
\end{eqnarray*}
\noindent This LPOD has four split programs:
\[
\begin{array}{l@{\hspace{5pt}}|@{\hspace{5pt}}l@{\hspace{5pt}}|@{\hspace{5pt}}l@{\hspace{5pt}}|@{\hspace{5pt}}l}
\begin{array}{rcl}
a & \leftarrow & \neg c\\
b & \leftarrow & \neg d
\end{array}
&
\begin{array}{rcl}
a & \leftarrow & \neg c\\
c & \leftarrow & \neg d \wedge \neg b
\end{array}
&
\begin{array}{rcl}
b & \leftarrow & \neg c \wedge \neg a\\
b & \leftarrow & \neg d
\end{array}
&
\begin{array}{rcl}
b & \leftarrow & \neg c \wedge \neg a\\
c & \leftarrow & \neg d \wedge \neg b
\end{array}
\end{array}
\]
\noindent that yield three answer sets $\{a,b\}$, $\{c\}$ and $\{b\}$.\qed
\end{example}

As explained in~\cite{BNS04}, answer sets of LPODs can also be described in terms of a program reduct, instead of using split programs. 
\begin{definition}[$\times$-reduct]
The $\times$-\emph{reduct} of an LPOD rule $r$ like \eqref{lpodrule} with respect to a set of atoms $I$ denoted as $r^I_\times$ and defined as the set of rules:
\begin{eqnarray}
\A[i] \leftarrow (\wedge \B) \label{xreduct}
\end{eqnarray}
\noindent for all $i=1,\dots,|\A|$ such that $I \models (\wedge \neg \B') \wedge (\wedge \neg \A[1..i\!-\!1]) \wedge \A[i]$.\qed
\end{definition}
\noindent As expected, the $\times$-\emph{reduct} of an LPOD $P$ with respect to $I$, written $P^I_\times$ is the union of all $r^I_\times$ for all LPOD rules $r \in P$. 
For instance, for $I=\{b,c\}$ and $P$: 
\begin{eqnarray}
a \times b & \leftarrow & c \wedge \neg d \label{f:ex2.1}\\
d \times a & \leftarrow & \neg b \label{f:ex2.2}\\
d \times e & \leftarrow & \neg a \label{f:ex2.3}
\end{eqnarray}
\noindent the reduct $P^I_\times$ would be the rule $\{b \leftarrow c\}$. Notice that $P^I_\times$ defined in this way is always a normal positive logic program and so it has a least model~\cite{vEK76}.

\begin{theorem}[From~\cite{BNS04}]
A set of atoms $I$ is an answer set of an LPOD $P$ iff $I \models P$ and $I$ is the least model of $P^I_\times$.\qed
\end{theorem}

It is important to note that $I \models P^I_\times$ does not imply $I \models P$, and thus, the latter is also required in the above theorem. For instance, in the last example, the interpretation $\emptyset$ is the least model of $P^I_\times$ but does not satisfy the LPOD rule \eqref{f:ex2.3}.




Although, as said in the introduction, we will concentrate on the answer sets of split programs, we include here the definition of the three ordering relations for selecting preferred answer sets among them. We say that an LPOD rule $r$ like \eqref{lpodrule} is \emph{satisfied to degree} $j \in \{1,\dots,|\A|\}$ by a set of atoms $I$, written $I \models_j r$, when: $I$ does not satisfy the body of $r$ and $j=1$; $I$ satisfies the body of $r$ and $j$ is the minimum index for which $\A[j] \in I$. We define $deg_I(r) \eqdef j$ when $I\models_j r$ and define the set $I^j(P) \eqdef \{r \in P \ | \ I\models_j r\}$. Given two answer sets $I, J$ of a given LPOD:
\begin{enumerate}
\item $I$ is \emph{cardinality-preferred} to $J$, written $I >_c J$, when for some truth degree $k$, $|I^k(P)|>|J^k(P)|$ whereas $|I^i(P)|=|J^i(P)|$ for all $i < k$.
\item $I$ is \emph{inclusion-preferred} to $J$, written $I >_i J$, when for some truth degree $k$, $I^k(P) \subset J^k(P)$ while $I^i(P)|=J^i(P)$ for all $i < k$.
\item $I$ is \emph{Pareto-preferred} to $J$, written $I >_p J$, if for some rule $r \in P$, $deg_I(r) < deg_J(r)$ whereas for no rule $r' \in P$, $deg_I(r')>deg_J(r')$.
\end{enumerate}

\section{The Ordered Disjunction Operator}\label{sec:od}

Let us consider the introduction of $\times$ in $\HT$ as the following derived operator:
\begin{eqnarray}
F \times G & \eqdef & F \vee (\neg F \wedge G) \label{f:x}
\end{eqnarray}


\noindent Although in classical logic, \eqref{f:x} $\equiv_c F \vee G$, this equivalence does not hold in $\HT$, that is $\eqref{f:x} \not\equiv_s F \vee G$. To see why, adding $G$, the two disjunctions have different equilibrium models: $\{F \vee G, G\}$ has one equilibrium model $\{G\}$, whereas $\{F \times G, G\}$ also has a second equilibrium model $\{F,G\}$. We discuss now some basic properties of $\times$ operator.
\begin{proposition}[Negation]\label{prop:neg}
The negation of an ordered and a regular disjunction are strongly equivalent:
\begin{eqnarray*}
\neg (F\times G) \ \equiv_s \ \neg F\wedge \neg G \ \equiv_s \ \neg (F\vee G)
\end{eqnarray*}
\end{proposition}
\begin{proof}
Applying De Morgan laws, $\neg (F\vee \neg F\wedge G)$ amounts to $\neg F\wedge (\neg \neg F\vee \neg G)$ which, by distributivity, is equivalent to $\neg F\wedge \neg \neg F\vee \neg F\wedge \neg G$, but the first disjunct can be removed, since $\neg F\wedge \neg \neg F$ is inconsistent in $\HT$ by \eqref{f:bot}.\qed
\end{proof}

\begin{proposition}[Truth constants]
These are some strongly equivalent simplifications:
\begin{eqnarray}
(F \times F) & \equiv_s & F \label{f:eq1}\\
(\bot \times F) & \equiv_s & F \label{f:eq2}\\
(F \times \bot) & \equiv_s & F \label{f:eq3}\\
(\top \times F) & \equiv_s & \top \label{f:eq4}\\
(F \times \top) & \equiv_s & (F \vee \neg F) \label{f:eq5}
\end{eqnarray}
\qed
\end{proposition}
\noindent Note that the main difference with respect to ordinary disjunction when dealing with truth constants is \eqref{f:eq5}. Equivalences \eqref{f:eq2} and \eqref{f:eq3} show that $\bot$ acts as a neutral element for ordered disjunction. This means that we can safely consider an empty ordered disjunction as $\bot$ (as happens with $\vee$ too).

Distributivity with respect to conjunction is satisfied in the following cases:
\begin{eqnarray*}
F \wedge (G \times H) & \equiv_s & (F \wedge G) \times (F \wedge H)\\
(F \times G) \wedge H & \equiv_s & (F \wedge H) \times (G \wedge H)\\
F \times (G \wedge H) & \equiv_s & (F \times G) \wedge (F \times H)
\end{eqnarray*}
\noindent but the following pair of formulas
\begin{eqnarray*}
(F \wedge G) \times H & \hspace{40pt} & (F \times H) \wedge (G \times H)
\end{eqnarray*}
\noindent are not strongly equivalent. In fact, they have different answer sets. The first rule has equilibrium models $\{F,G\}$ and $\{H\}$ whereas the second has two additional ones $\{F,H\}$ and $\{G,H\}$.
Distributivity between disjunctions only holds in the following case:
\begin{eqnarray*}
F \times (G \vee H) \equiv_s (F \times G) \vee (F \times H)
\end{eqnarray*}
\noindent but the following list shows pairs of formulas that are \emph{not} strongly equivalent:
\begin{eqnarray}
F \vee (G \times H) & \hspace{40pt} & (F \vee G) \times (F \vee H) \label{f:a1} \\
(F \times G) \vee H & \hspace{40pt} & (F \vee H) \times (G \vee H) \label{f:a2}\\
(F \vee G) \times H & \hspace{40pt} & (F \times H) \vee (G \times H) \label{f:a3}
\end{eqnarray}
\noindent Take line \eqref{f:a1}, for instance. Adding the atom $H$, the left rule yields two equilibrium models $\{G,H\}$ and $\{H\}$, whereas the second rule allows a third equilibrium model $\{F,H\}$. Intuitively, adding $H$ to $F \vee (G \times H)$ will always make the second disjunct to be true (both when $G$ holds and when it does not). So, there is no need to make $F$ true.
By a similar reason, in the case of line \eqref{f:a2}, after adding atom $G$, the first rule yields two equilibrium models, $\{F,G\}$ and $\{G\}$ (there is no need to make $H$ true), while the second rule yields a third solution $\{G,H\}$. Finally, for line \eqref{f:a3}, if we add the atom $H$ to the first expression, we get three equilibrium models, $\{F,H\}$, $\{G,H\}$, and $\{H\}$, whereas for the second expression, the addition of $H$ only yields the equilibrium model $\{H\}$ itself. 


\begin{proposition}[Associativity]\label{prop:asoc}
The $\times$ operator is associative, that is:
\begin{eqnarray*}
F \times (G \times H) & \equiv_s & (F \times G) \times H
\end{eqnarray*}
\end{proposition}
\begin{proof}
\[
\begin{array}{rcl@{\hspace{10pt}}l}
F \times (G \times H) & \equiv_s & F \vee \neg F \wedge (G \vee \neg G \wedge H)
& \hbox{By definition of } \times.\\
& \equiv_s & F \vee \neg F \wedge G \vee \neg F \wedge \neg G \wedge H & \hbox{Distribuitivity.}\\
& \equiv_s & (F \times G) \vee \neg F \wedge \neg G \wedge H 
& \hbox{Definition of } \times.\\
& \equiv_s & (F \times G) \vee \neg (F \times G) \wedge H 
& \hbox{By Proposition~\ref{prop:neg}.}\\
& \equiv_s & (F \times G) \times H & \hbox{Definition of } \times.
\end{array}
\]\qed
\end{proof}

The third line of the proof suggests the following translation for a sequence of consecutive applications of $\times$.

\begin{theorem}\label{th:unfold1}
Let $\A$ be a sequence of $|\A|=n \geq 0$ formulas. Then, the formula $(\times \A)$ is strongly equivalent to:
\begin{eqnarray}
\bigvee_{i=1,\dots, n} (\wedge \neg \A[1..i\!\!-\!\!1]) \wedge \A[i] \label{f:t1}
\end{eqnarray}
\end{theorem}
\begin{proof}
By induction on $n=|\A|$. When $n=0$, both $(\times \A)$ and \eqref{f:t1} correspond to $\bot$. Assume it is proved for any sequence of length $n-1$, with $n>0$. Since $\times$ is associative, we can write $(\times \A)$ as $(\times \A[1..n\!-\!1]) \times \A[n]$. Let us call $\alpha$ to $(\times \A[1..n\!-\!1])$. Now, by definition of $\times$, $(\times \A)$ corresponds to $\alpha \vee \neg \alpha \wedge \A[n]$. On the other hand, by induction, $\alpha$ is equivalent to:
\begin{eqnarray*}
\bigvee_{i=1,\dots, n-1} (\wedge \neg \A[1..i\!\!-\!\!1]) \wedge \A[i]
\end{eqnarray*}
\noindent whereas $\neg \alpha$ is equivalent to $(\wedge \neg \A[1..n\!-\!1])$, by Proposition~\ref{prop:neg}. But then, $\alpha \vee \neg \alpha \wedge \A[n]$ corresponds to:
\begin{eqnarray*}
\underbrace{\bigg[ \ \bigvee_{i=1,\dots, n-1} (\wedge \neg \A[1..i\!\!-\!\!1]) \wedge \A[i] \ \bigg]}_{\alpha}\ \vee \ \underbrace{(\wedge \neg \A[1..n\!-\!1])}_{\neg \alpha} \wedge \A[n]
\end{eqnarray*}
\noindent which is exactly the expansion in \eqref{f:t1}.\qed
\end{proof}
As an example, the expression $a \times b \times c \times d$ is strongly equivalent to:
\begin{eqnarray*}
a \vee (\neg a \wedge b) \vee (\neg a \wedge \neg b \wedge c) \vee (\neg a \wedge \neg b \wedge \neg c \wedge d).
\end{eqnarray*}

Although we have already proved that $\times$ satisfies the idempotence property \eqref{f:eq1}, it must be noticed that since the operator is obviously non-commutative, a repeated subformula in a sequence of ordered disjunctions cannot always be removed. We can use Theorem~\ref{th:unfold1} to prove, for instance, the following property.
\begin{proposition}[Ordered Idempotence]
$(F \times G \times F) \equiv_s F \times G$.
\end{proposition}
\begin{proof}
Just notice that, by Theorem~\ref{th:unfold1}, $(F \times G \times F)$ is equivalent to $F \vee (\neg F \wedge G) \vee (\neg F \wedge \neg G \wedge F)$, but the last disjunct can be removed due to \eqref{f:bot}.\qed
\end{proof}

\noindent As $F$ and $G$ are arbitrary formulas, and $\times$ is associative, the equivalence above implies that, when repeated atoms occur in an ordered disjunction, all their occurrences can be removed, excepting the leftmost one. For instance:
\begin{eqnarray*}
a \times b \times c \times a \times d \times c \times a \times e \times b & \leftrightarrow & a \times b \times c \times d \times e
\end{eqnarray*}
However, in the general case, we cannot remove the leftmost occurrence of a repeated formula. As a counterexample, the formulas $a \times b \times a$ and $b\times a$ are not strongly equivalent: the simple addition of atom $a$ to the first formula yields one equilibrium model $\{a\}$ whereas, when added to the second formula, we also obtain an additional equilibrium model $\{a,b\}$. 

The following second translation is perhaps more interesting both for generating a logic program, and for comparison purposes with the $\times$-reduct. We will introduce first a useful result.

\begin{theorem}\label{th:unfold2}
Let $\A$ be a sequence of $|\A|=n \geq 0$ formulas. Then, the formula $(\times \A)$ is strongly equivalent to the conjunction of:
\begin{eqnarray}
\bigwedge_{i=1,\dots,n} \bigg( \A[i] \vee \neg \A[i] & \leftarrow & (\wedge \neg \A[1..i\!-\!1]) \ \bigg)\label{f:t2.1} \\
\bot & \leftarrow & (\wedge \neg \A) \label{f:t2.2}
\end{eqnarray}
\end{theorem}
\begin{proof}
First, note that both formulas are classically equivalent: all conjuncts in \eqref{f:t2.1} for $i=1,\dots,n$ are classical tautologies whereas \eqref{f:t2.2} $\equiv_c (\vee \A) \equiv_c (\times \A)$. 

We prove now that $\tuple{H,T} \models (\times \A)$ iff $\tuple{H,T} \models \eqref{f:t2.1}\wedge \eqref{f:t2.2}$. For the left to right direction, we begin showing that $(\times \A)$ implies $\eqref{f:t2.1}\wedge \eqref{f:t2.2}$. Assume we had some $\tuple{H,T} \models (\times \A)$ but $\tuple{H,T} \not\models \A[i] \vee \neg \A[i] \leftarrow (\wedge \neg \A[1..i\!-\!1])$ for some $i=1,\dots, n$. As these implications are classical tautologies, the only possibility is $\tuple{H,T} \models (\wedge \neg \A[1..i\!-\!1])$ while $\tuple{H,T} \not\models \A[i] \vee \neg \A[i]$. We will prove that, if so, $\HTtup \not\models \eqref{f:t1}$ which, by Theorem~\ref{th:unfold1}, means $\HTtup \not\models (\times \A)$, reaching a contradiction. Notice that we have $\tuple{H,T} \models \neg \A[1], \dots, \tuple{H,T} \models \neg \A[i\!-1\!]$ while $\tuple{H,T} \not\models \A[i]$ and $\tuple{H,T} \not\models \neg \A[i]$. Now, the first $i\!-\!1$ disjuncts in \eqref{f:t1} will be false because for all $j=1,\dots,i\!-\!1$, $\tuple{H,T} \models \neg \A[j]$ implies $\tuple{H,T} \not\models \A[j]$. The $i$-th disjunct is false because it contains $\A[i]$ and $\tuple{H,T} \not\models \A[i]$, and the rest of disjuncts are also false since they contain the subformula $\neg \A[i]$ and we also had $\tuple{H,T} \not\models \neg \A[i]$.

We remain to prove that $\tuple{H,T} \models \eqref{f:t2.2}$. By Proposition~\ref{prop:neg}, $(\wedge \neg \A)$ is equivalent to $\neg (\times \A)$ and thus, \eqref{f:t2.2} corresponds to $\neg \neg (\times \A)$. But in $\HT$  (and in fact, in intuitionistic logic), any formula implies its double negation, and we had $\tuple{H,T} \models (\times \A)$.

For the right to left direction, suppose $\tuple{H,T} \models \eqref{f:t2.1}\wedge\eqref{f:t2.2}$ but $\tuple{H,T} \not\models (\times \A)$. From the latter and Theorem~\ref{th:unfold1} we conclude the following condition, let us call it (a): for all $i=1,\dots,n$, $\tuple{H,T} \not\models \A[i]$ or there exists some $j$, $1\leq j < i$ such that $\tuple{H,T} \not\models \neg \A[j]$. We prove now, by induction on $i$ in \eqref{f:t2.1}, that $\HTtup \models \neg \A[i]$ for all $i=1,\dots,n$, which contradicts $\HTtup \models \eqref{f:t2.2}$. For $i=1$ there is no $j$ smaller and from (a) we get $\HTtup \not\models \A[1]$. But, with $i=1$, the antecedent of \eqref{f:t2.1} becomes empty ($\top$) and we get $\HTtup \models \A[1] \vee \neg \A[1]$ that, as $\HTtup \not\models \A[1]$, leads to $\HTtup \models \neg \A[1]$. Induction hypothesis: suppose we have proved $\HTtup \models \neg \A[j]$ for all $j$, $1 \leq j < i$. From this together with (a) we conclude $\HTtup \not\models \A[i]$. But at the same time, this induction hypothesis means that $\HTtup$ satisfies the antecedent of \eqref{f:t2.1}, so we conclude $\HTtup \models \A[i] \vee \neg \A[i]$, but again, as $\HTtup \not\models \A[i]$ the only possibility is $\HTtup \models \neg \A[i]$.\qed
\end{proof}

The previous theorem can be combined with the following lemma to obtain a translation of any LPOD into a logic program.

\begin{lemma}\label{lem:b}
If $\varphi$ is strongly equivalent to a conjunction of implications $(\alpha_i \rightarrow \beta_i)$ with $i=1,\dots,n$ and $n \geq 0$ then the formula $F: A \rightarrow \varphi \vee B$ is strongly equivalent to $G$, the conjunction of $\alpha_i \wedge A \rightarrow \beta_i \vee B$.\end{lemma}
\begin{proof}
Take $\varphi$ strongly equivalent to the conjunction of $(\alpha_i \rightarrow \beta_i)$. To prove that $F \equiv_s G$ we must show both that $F \equiv_c G$; and that $\HTtup \models F$ iff $\HTtup \models G$.

For proving the classical equivalence, notice that $F \equiv_c \neg A \vee \varphi \vee B$. But as strong equivalence implies classical equivalence, we can replace $\varphi$ in the last formula by the conjunction of implications, obtaining $\neg A \vee B \vee \bigwedge_i (\alpha_i \rightarrow \beta_i) \equiv_c \bigwedge \neg A \vee B \vee (\alpha_i \rightarrow \beta_i) \equiv_c G$.

For proving $\HTtup \models F$ iff $\HTtup \models G$, assume as a first case that $\HTtup \not\models A$ or $\HTtup \models B$. If so, we trivially have both $\HTtup \models F$ and $\HTtup \models G$, and the equivalence holds. Suppose, on the contrary, that $\HTtup \models A$ and $\HTtup \not\models B$. It is easy to see that, then, $\HTtup \models F$ iff $\HTtup \models \varphi$. Similarly $\HTtup \models G$ amounts to $\HTtup \models \bigwedge_i (\alpha_i \rightarrow \beta_i)$ and the latter, due to our premise, is equivalent to $\HTtup \models F$.\qed
\end{proof}

In particular, we can thus translate any rule like \eqref{lpodrule} by using Lemma~\ref{lem:b} with $H=(\wedge \B) \wedge (\wedge \neg \B')$, $G=\bot$ and $F=(\times \A)$ and Theorem~\ref{th:unfold1} to obtain a conjunction of implications equivalent to $F$. In this way, for any LPOD rule $r$ like \eqref{lpodrule} we define the set of rules $r^*$ as the conjunction of implications in \eqref{f:t2.1} and \eqref{f:t2.2} after adding to all their bodies the conjunction of $(\wedge \B) \wedge (\wedge \neg \B')$, that is:
\begin{eqnarray}
\A[i] \vee \neg \A[i] & \leftarrow & (\wedge \B) \wedge (\wedge \neg \B') \wedge (\wedge \neg \A[1..i\!-\!1])   \label{f:t3.1}\\
\bot & \leftarrow & (\wedge \B) \wedge (\wedge \neg \B') \wedge (\wedge \neg \A) \label{f:t3.2}
\end{eqnarray}
\noindent for all $i=1, \dots, |\A|$. For any LPOD $P$, the program $P^*$ stands for the union of all $r^*$ for each $r\in P$. As an example, given the LPOD rule $r: a \times b \times c \leftarrow p \wedge \neg q$, the set of rules $r^*$ consists\footnote{In fact, this translation can be further refined by removing the last constraint and the last negative literal in the head. In the example, we would replace the last two rules by $c \leftarrow p \wedge \neg q \wedge \neg a \wedge \neg b$.}
 of:
\begin{eqnarray*}
a \vee \neg a & \leftarrow & p \wedge \neg q \\
b \vee \neg b & \leftarrow & p \wedge \neg q \wedge \neg a\\
c \vee \neg c & \leftarrow & p \wedge \neg q \wedge \neg a \wedge \neg b\\
         \bot & \leftarrow & p \wedge \neg q \wedge \neg a \wedge \neg b \wedge \neg c
\end{eqnarray*}

\begin{theorem}\label{th:pstar}
For any LPOD $P$, $P \equiv_s P^*$.
\end{theorem}
\begin{proof}
It directly follows from Theorem~\ref{th:unfold2} and Lemma~\ref{lem:b}.\qed
\end{proof}

An important remark is that, although $P^*$ is a disjunctive logic program with negation in the head, it belongs to a subclass of disjunctive programs with the same expressiveness and complexity than those of normal programs. To see why, it suffices to observe that any formula of the form $F \vee \neg p \leftarrow G$ is strongly equivalent to $F \leftarrow G \wedge \neg \neg p$ and in its turn, a double negated atom in the body can be replaced\footnote{Similar techniques for removing head negations were introduced in~\cite{IS98,Jan01}.} by the negation of a fresh auxiliary atom $F \leftarrow G \wedge \neg aux$ plus the rule $aux \leftarrow \neg p$.

\begin{lemma}\label{lem:1}
Let $I$ and $J$ be sets of atoms and $P$ an LPOD. Then ($J \models P^I_\times$ and $I \models P$) iff $J \models (P^*)^I$.
\end{lemma}
\begin{proof}
Take any rule $r \in P$ of the form \eqref{lpodrule}. If we apply the standard reduct with respect to $I$ on the rules like \eqref{f:t3.1} in $r^*$, it is easy to see that we exactly get the rules in $r^I_\times$. Thus, $P^I_\times$ coincides with the standard reduct for all the rules like \eqref{f:t3.1} in $P^*$ with respect to $I$. Now, let us call $Q$ to set of remaining rules in $P^*$ of the form \eqref{f:t3.2}. We will prove that $I \models P$ iff $J \models Q^I$. In fact, for each $r \in P$ like \eqref{lpodrule} we have a corresponding $r' \in Q$ like \eqref{f:t3.2} and we show that $I \models r$ iff $J \models (r')^I$. We begin observing that $r'$ \eqref{f:t3.2} and $r$ \eqref{lpodrule} are classically equivalent formulas. We have then two cases: if $I \models (\wedge \B) \wedge (\wedge \neg \B') \wedge (\wedge \neg \A)$ then $r^I$ is $\bot$ and trivially $J \not\models \bot$, but also clearly $I \not\models r'$ and thus $I \not\models r$, and the equivalence holds. If $I \not\models (\wedge \B) \wedge (\wedge \neg \B') \wedge (\wedge \neg \A)$ the reduct $(r')^I$ is empty and thus $J \models (r')^I$ trivially, but we also have $I \models r'$ and so $I \models r$ and again the equivalence holds.\qed
\end{proof}

\begin{lemma}\label{lem:2}
$I$ is an answer set of an LPOD $P$ iff $I$ is an answer set of $P^*$.\qed
\end{lemma}
\begin{proof}
For the left to right direction, suppose $I$ is an answer set of $P$. Then, $I \models P$ and $I$ is a minimal model of $P^I_\times$. By Lemma~\ref{lem:1}, this implies $I \models (P^*)^I$. Suppose there exists a smaller $J \subset I$ such that $J \models (P^*)^I$. Applying Lemma~\ref{lem:1} again, we conclude that $J \models P^I_\times$ and this contradicts that $I$ is minimal model of $P^I_\times$.

For the right to left direction, the reasoning is analogous. If $I$ is an answer set of $P^*$ we obviously have $I \models (P^*)^I$ and, applying Lemma~\ref{lem:1} we conclude $I \models P$ and $I \models P^I_\times$. Assume we have some $J \subset I$, $J \models P^I_\times$. Again, by Lemma~\ref{lem:1} we conclude $J \models (P^*)^I$ and we get a contradiction with minimality of $I$ as model of $(P^*)^I$.\qed
\end{proof}

\begin{theorem}[main theorem]\label{th:LPOD}
$T$ is an answer set of an LPOD $P$ iff $\tuple{T,T}$ is an equilibrium model of $P$.
\end{theorem}
\begin{proof}
It directly follows from Lemma~\ref{lem:2}, Theorem~\ref{th:pstar} and Theorem~\ref{th:as}.\qed
\end{proof}

As a final comment, a third possible representation of $(\times \A)$ is the strongly equivalent formula:
\begin{eqnarray*}
(\vee \A) & \wedge & \bigwedge_{i=1,\dots,n-1} \big( \ (\vee \A[1..i]) \vee \neg \A[i] \ \big)
\end{eqnarray*}
For instance, $a \times b \times c$ is strongly equivalent to:
\begin{eqnarray*}
(a \vee b \vee c) \wedge (a \vee \neg a) \wedge (a \vee b \vee \neg b) \wedge (a \vee b \vee c \vee \neg c).
\end{eqnarray*}

\section{Disjunctive LPOD}\label{sec:dlpod}

In~\cite{KLOP08} an extension of LPODs for dealing with regular disjunction $\vee$ is considered. An \emph{ordered disjunctive term} is defined as any arbitrary combination of atoms and $\vee, \times$ operators. A \emph{Disjunctive LPOD} (DLPOD) is a set of rules of the form $F \leftarrow (\wedge \B) \wedge (\wedge \neg \B')$ where $F$ is an ordered disjunctive term and $\B, \B'$ are lists of atoms. An ordered disjunctive term is said to be in \emph{Ordered Disjunctive Normal Form} (ODNF) if it has the form $(\times \A_1) \vee \dots \vee (\times \A_n)$ where the $\A_i$ are lists of atoms. A DLPOD rule is in ODNF if its head is in ODNF. Similarly, the whole DLPOD is in ODNF if all its rules are in ODNF. In fact, the semantics of an arbitrary DLPOD is defined by first translating its head into ODNF using the following 
rewriting rules:
\begin{eqnarray}
F \times (G \vee H) & \longmapsto & (F \times G) \vee (F \times H) \label{f:rw1} \\
(F \vee G) 	\times H & \longmapsto & (F \times H) \vee (G \times H) \label{f:rw2}\\
(F \times G) \times H & \longmapsto & F \times G \times H \label{f:rw3}\\
F \times (G \times H) & \longmapsto & F \times G \times H \label{f:rw4}
\end{eqnarray}
\noindent The exhaustive application of \eqref{f:rw1}-\eqref{f:rw4} allow transforming any DLPOD into ODNF, that is, a set of rules like
\begin{eqnarray}
\bigvee_{i=1,\dots,n} (\times \A_i) \leftarrow (\wedge \B) \wedge (\wedge \neg \B') \label{dlpodrule}
\end{eqnarray}
\noindent where $n \geq 0$, and $\B, \B'$ and all the $\A_i$ are lists of atoms. For any DLPOD rule $r$ like \eqref{dlpodrule}, an \emph{option} for $r$ is any rule of the form:
\begin{eqnarray}
\bigvee_{i=1,\dots,n} \A_i[k_i] \leftarrow  (\wedge \B) \wedge (\wedge \neg \B') \wedge \bigwedge_{i=1,\dots,n} (\wedge \neg \A_i[1..k_i\!\!-\!\!1] ) \label{dlpodopt}
\end{eqnarray}
\noindent where each $k_i$ is some value $k_i \in \{1,\dots,|\A_i|\}$. For instance, the options  for the DLPOD rule $(a \times b) \vee (c \times d \times e)$ are the rules:
\[
\begin{array}{l@{\hspace{20pt}}l}
\begin{array}{rcl}
a \vee c & & \\
a \vee d & \leftarrow & \neg c\\
a \vee e & \leftarrow & \neg c \wedge \neg d
\end{array}
&
\begin{array}{rcl}
b \vee c & \leftarrow & \neg a\\
b \vee d & \leftarrow & \neg a \wedge \neg c\\
b \vee e & \leftarrow & \neg a \wedge \neg c \wedge \neg d
\end{array}
\end{array}
\]

A disjunctive logic program $P'$ is a \emph{split program} of a DLPOD $P$ if it is the result of replacing each DLPOD rule $r \in P$ by some of its split rules. It is important to notice that the split programs of a DLPOD are disjunctive logic programs, whereas in the case of LPODs, split programs were normal logic programs. As before, a set of atoms $I$ is an \emph{answer set} of a DLPOD $P$ if it is an answer set of some split program $P'$ of $P$. For comparison purposes, we concentrate here on the answer sets of DLPODs, leaving apart the selection of their preferred answer sets (see~\cite{KLOP08} for further details).

\subsection{Comparison}

Since our characterisation of $\times$ applies for any arbitrary propositional theory, in particular, it also provides a semantics for DLPODs, both arbitrary and in ODNF. So the immediate question is whether our $\HT$ based characterisation and the original DLPOD semantics coincide. Unfortunately, the answer to this question is negative, by two reasons. First, as we explained in Section~\ref{sec:od} with a counterexample for \eqref{f:a1}, we saw that in fact, \eqref{f:rw2} is not an strongly equivalent transformation. Thus, under our $\HT$-based characterisation, an arbitrary DLPOD is not always reducible to ODNF (although, of course, a semantics for the DLPOD is still defined). Still, even when we restrict ourselves to the study of ODNF programs, we can find examples where the original DLPOD semantics provides more answer sets.

\begin{example}\label{ex2}
Take the DLPOD $P_{\ref{ex2}}$ consisting of the pair of rules
\begin{eqnarray*}
a \vee (b \times c) & \hspace{20pt} & c
\end{eqnarray*}
The split programs of $P_{\ref{ex2}}$ are $\{(a \vee b),c\}$ and $\{(a \vee c \leftarrow \neg b),c\}$, the first one with two answer sets $\{a,c\}$, $\{b,c\}$ and the second one with answer set $\{c\}$. In this case, there is one answer set that is \emph{not} equilibrium model: $\{a,c\}$.\qed
\end{example}

Answer set $\{a,c\}$ looks counterintuitive in the sense that, once $c$ is true and we are not picking choice $b$, it seems clear that $(b \times c)$ holds and there is no clear reason to make $a$ true (in the first rule). Remember that $\vee$ is a regular disjunction, and it will become satisfied when any of its disjuncts become true, regardless the truth degree.
The next example illustrates a counterintuitive effect of DLPOD semantics when defining subformulas with auxiliary atoms.

\begin{example}\label{ex3}
Let $P_{\ref{ex3}}$ be the DLPOD consisting of the pair of rules:
\begin{quote}
\hspace{60pt} $a \times (b \vee c)  \hspace{40pt}  c$ \qed
\end{quote}
\end{example}
\noindent For both semantics, $P_{\ref{ex3}}$ can be converted (even under strong equivalence in $\HT$) into the ODNF program $P'_{\ref{ex3}}$ consisting of rule $(a \times b) \vee (a \times c)$ and fact $c$. Under the $\HT$ characterisation, both $P_{\ref{ex3}}$ and $P'_{\ref{ex3}}$ have two answer sets, $\{a,c\}$ and $\{c\}$. For DLPOD semantics, $P'_{\ref{ex3}}$  has a third, additional answer set $\{b, c\}$. Now, notice that a different way of computing the answer sets of the original program $P_{\ref{ex3}}$ could be defining $b \vee c$ with an auxiliary atom, so we obtain the LPOD $P''_{\ref{ex3}}$:
\[
\begin{array}{c@{\hspace{20pt}}rcl@{\hspace{20pt}}rcl}
a \times aux    & aux & \leftarrow & b & b \vee c & \leftarrow & aux\\
c               & aux & \leftarrow & c
\end{array}
\]
\noindent It is easy to see that $P''_{\ref{ex3}}$ has two answer sets $\{aux, a, c\}$ and $\{aux, c\}$ which, after removing the auxiliary atom, become $\{a,c\}$ and $\{c\}$, respectively, which were, in fact, the two answer sets we obtained for $P_{\ref{ex3}}$ using the $\HT$ characterisation. This was expected, since $\HT$ satisfies the rule of substitution, that is, if $aux \leftrightarrow \alpha$ we can always replace $\alpha$ by $aux$. In these counterexamples, DLPOD semantics yielded more answer sets than the $\HT$ characterisation -- this is actually a general property stated below\footnote{We leave the proof for a future extended version of this document.}.
\vspace{-5pt}
\begin{theorem}
If $\tuple{T,T}$ is an equilibrium model of a DLPOD $P$ in ODNF, then $T$ an answer set of $P$.\qed
\end{theorem}

\section{Conclusions}\label{sec:conc}

We have presented a logical characterisation of Logic Programs with Ordered Disjunction (LPOD) that allows a direct study of ordered disjunction $\times$ as a derived operator in the logic of Here-and-There ($\HT$), well-known for its application to (strongly equivalent) logic program transformations in Answer Set Programming. This characterisation provides an alternative implementation of LPODs that does not resort to auxiliary predicates. It has also allowed us to analyse the behavior of the $\times$ operator with respect to some typical properties like associativity, distributivity, idempotence, etc. As $\times$ is handled as a regular logical connective, our characterisation covers any arbitrary syntactic extension, and in particular, the so-called Disjunctive LPOD (DLPOD). We have shown that the semantics of DLPODs shows some differences with respect to the $\HT$ characterisation and established a formal comparison.

our result can also be seen as a confirmation of Theorem 6 in~\cite{FTW08}. In that work, a reduct-based formalisation of LPODs was proposed in order to study strong equivalence relations among LPODs and regular programs. Theorem 6 in that work showed that their characterisation of strong equivalence for LPODs actually coincided with the one for regular programs. This result becomes trivial under our current approach, since ordered disjunction is just treated as an $\HT$ derived operator and, as proved in~\cite{CF07}, $\HT$ arbitrary theories are strongly equivalent to logic programs.

We have implemented a first prototype for propositional LPODs using the current approach that uses {\tt DLV} system as a backend\footnote{See \url{http://www.dc.fi.udc.es/~cabalar/lpod/}}. Future work includes the extension of this prototype to deal with variables and with arbitrary combinations of ordered and regular disjunction in the head.

\bibliographystyle{plain}
\bibliography{od}

\begin{thebibliography}{10}

\bibitem{BM03}
Marcello Balduccini and Veena~S. Mellarkod.
\newblock {CR}-{P}rolog with ordered disjunction.
\newblock In {\em Proceedings of the 2nd Intl. on Answer Set Programming,
  Advances in Theory and Implementation (ASP'03)}, 2003.

\bibitem{BMP05}
Elisa Bertino, Alessandra Mileo, and Alessandro Provetti.
\newblock {PDL} with preferences.
\newblock In {\em Proc. of the 6th IEEE International Workshop on Policies for
  Distributed Systems and Networks (POLICY 2005)}, pages 213--222, 2005.

\bibitem{BNS04}
Gerhard Brewka, Ilkka Niemel{\"a}, and Tommi Syrj{\"a}nen.
\newblock Logic programs with ordered disjunction.
\newblock {\em Computational Intelligence}, 20(2):335--357, 2004.

\bibitem{CF07}
P.~Cabalar and P.~Ferraris.
\newblock Propositional theories are strongly equivalent to logic programs.
\newblock {\em Theory and Practice of Logic Programming}, 7(6):745--759, 2007.

\bibitem{DSTW04}
J.~P. Delgrande, T.~Schaub, H.~Tompits, and K.~Wang.
\newblock A classification and survey of preference handling approaches in
  nonmonotonic reasoning.
\newblock {\em Computational Intelligence}, 20(2):308--334, 2004.

\bibitem{FTW08}
Wolfgang Faber, Hans Tompits, and Stefan Woltran.
\newblock Notions of strong equivalence for logic programs with ordered
  disjunction.
\newblock In Gerhard Brewka and J{\'e}r{\^o}me Lang, editors, {\em Proc. of the
  11th Intl. Conf. on Principles of Knowledge Representation and Reasoning
  (KR'08)}, pages 433--443. AAAI Press, 2008.

\bibitem{FLL07}
P.~Ferraris, J.~Lee, and V.~Lifschitz.
\newblock A new perspective on stable models.
\newblock In {\em Proc. of the Intl. Joint Conf. on Artificial Intelligence
  (IJCAI'07)}, 2007.

\bibitem{Fer05}
Paolo Ferraris.
\newblock Answer sets for propositional theories.
\newblock In {\em Proc. of the 8th Intl. Conf. on Logic Programming and
  Nonmonotonic Reasoning}, 2005.
\newblock (to appear).

\bibitem{FMB04}
Norman~Y. Foo, Thomas Meyer, and Gerhard Brewka.
\newblock {LPOD} answer sets and {N}ash equilibria.
\newblock In {\em Proc. of 9th Asian Computing Science Conference (ASIAN'04)},
  pages 343--351, 2004.

\bibitem{GL88}
Michael Gelfond and Vladimir Lifschitz.
\newblock The stable models semantics for logic programming.
\newblock In {\em Proc. of the 5th Intl. Conf. on Logic Programming}, pages
  1070--1080, 1988.

\bibitem{Hey30}
Arend Heyting.
\newblock Die formalen {R}egeln der intuitionistischen {L}ogik.
\newblock {\em Sitzungsberichte der Preussischen Akademie der Wissenschaften,
  Physikalisch-mathematische Klasse}, pages 42--56, 1930.

\bibitem{IS98}
Katsumi Inoue and Chiaki Sakama.
\newblock Negation as failure in the head.
\newblock {\em Journal of Logic Programming}, 35(1):39--78, 1998.

\bibitem{Jan01}
Tommi Janhunen.
\newblock On the effect of default negation on the expressiveness of
  disjunctive rules.
\newblock In {\em Proc. of the 6th Intl. Conf. on Logic Programming and
  Nonmonotonic Reasoning (LPNMR'01)}, pages 93--106, 2001.

\bibitem{KLOP08}
Philipp K{\"a}rger, Nuno Lopes, Daniel Olmedilla, and Axel Polleres.
\newblock Towards logic programs with ordered and unordered disjunction.
\newblock In {\em Workshop on Answer Set Programming and Other Computing
  Paradigms (ASPOCP 2008), 24th International Conference on Logic Programming
  (ICLP 2008)}, Udine, Italy, 12 2008.

\bibitem{LPV01}
Vladimir Lifschitz, David Pearce, and Agust\'{\i}n Valverde.
\newblock Strongly equivalent logic programs.
\newblock {\em ACM Transactions on Computational Logic}, 2:526--541, 2001.

\bibitem{MMP05}
Massimo Marchi, Alessandra Mileo, and Alessandro Provetti.
\newblock Declarative policies for web service selection.
\newblock In {\em Proc. of the 6th IEEE International Workshop on Policies for
  Distributed Systems and Networks (POLICY 2005)}, pages 239--242, 2005.

\bibitem{MT99}
V.~Marek and M.~Truszczy{\'n}ski.
\newblock {\em Stable models and an alternative logic programming paradigm},
  pages 169--181.
\newblock 1999.

\bibitem{Nie99}
I.~Niemel{\"a}.
\newblock Logic programs with stable model semantics as a constraint
  programming paradigm.
\newblock {\em Annals of Mathematics and Artificial Intelligence}, 25:241--273,
  1999.

\bibitem{Pea97}
David Pearce.
\newblock A new logical characterisation of stable models and answer sets.
\newblock In {\em Non monotonic extensions of logic programming. Proc.
  NMELP'96. (LNAI 1216)}. Springer-Verlag, 1997.

\bibitem{vEK76}
M.~H. van Emden and R.~A. Kowalski.
\newblock The semantics of predicate logic as a programming language.
\newblock {\em Journal of the ACM}, 23:733--742, 1976.

\bibitem{ZONSS05}
Claudia Zepeda, Mauricio Osorio, Juan~Carlos Nieves, Christine Solnon, and
  David Sol.
\newblock Applications of preferences using answer set programming.
\newblock In {\em In Proceedings Workshop on Answer Set Programming (ASPÕ05)},
  pages 318--332, 2005.

\end{thebibliography}
\end{document}